\documentclass[aps,prl,twocolumn,superscriptaddress,showpacs,floatfix]{revtex4-2}
\usepackage{graphicx,float,bm,epsfig,textcomp,color,dcolumn,setspace,array,calrsfs,mathrsfs,amsmath,amssymb,mathptmx,gensymb,booktabs,url,bibentry,natbib,subfigure,float,braket,lipsum,xcolor,relsize}
\usepackage[T1]{fontenc}
\usepackage[mathscr]{eucal}
\usepackage[bookmarks=true,colorlinks,linkcolor=blue,urlcolor=blue,citecolor=blue,plainpages=false,pdfpagelabels,final,breaklinks=true]{hyperref}
\usepackage{cleveref}
\usepackage{tikz}

\DeclareMathAlphabet{\mathcal}{OMS}{cmsy}{m}{n}
\makeatletter
\newcommand{\supplementarytableofcontents}{%
  \@starttoc{stoc}%
}
\makeatother

\begin{document}
\title{Phononic enhancement and detection of hidden spin-nematicity and dynamics in quantum magnets}
\author{Junyu Tang}
\affiliation{International Center for Quantum Materials, School of Physics, Peking University, Beijing 100871, China}
\author{Hong-Hao Song}
\affiliation{International Center for Quantum Materials, School of Physics, Peking University, Beijing 100871, China}
\author{Gang v.~Chen}
\email{chenxray@pku.edu.cn}
\affiliation{International Center for Quantum Materials, School of Physics, Peking University, Beijing 100871, China}
\affiliation{Collaborative Innovation Center of Quantum Matter, 100871, Beijing, China}

\begin{abstract}
The spin nematic phase, characterized by long-range order of spin quadrupole moments in the absence 
of dipolar magnetism, presents a significant challenge for conventional experimental detection. 
We propose a novel method to detect this elusive order in quantum magnets with an illustration 
in the spin-1 triangular lattice Mott insulator. By integrating out the phonon degrees of freedom, 
we obtain a phase diagram with substantially enlarged regions for the spin-nematic and 
spin-nematic-supersolid phases. We then demonstrate that through the spin-lattice coupling, 
the emergence of spin nematic order imprints a distinctive signature onto the phonon spectra, 
providing a clear spectroscopic signature for the quadrupolar order accessible via Raman 
or inelastic X-ray scattering. Our formalism offers a direct and powerful method to uncover the
hidden spin nematicity, opening a new pathway for diagnosing multipolar orders in quantum magnets.
\end{abstract}


\maketitle

Quantum magnets with the spin moments greater than 1/2 can host exotic phases of matter that extend beyond traditional magnetic order~\cite{Penc2011,Pourovskii2025}. For higher spin systems, the local moments are characterized not only by a dipole moment but also by the multipole moment such as the quadrupole moment~\cite{Tsunetsugu_2006_JPSJ,PAPANICOLAOU1988367,Zhentao_2025_PRL}. The ordering of these quadrupole moments leads to a spin nematic phase~\cite{Bhattacharjee_2006_PRB, Starykh_2015,Coleman_1991_PRL,Shannon_2013_PRB,huang2026}, a quantum paramagnetic state analogous to the liquid crystals, where the conventional magnetic dipolar order is absent but the spin system spontaneously breaks the rotational symmetry in the spin space.


A major obstacle in the study of the spin nematics is the difficulty in detecting them experimentally~\cite{Podolsky_2005,Yoshimitsu_2019_PNAS,Kim2024}. Since the neutron scattering primarily couples to the magnetic dipole moments~\cite{Halpern_1939_PR,Brockhouse_1957_PR}, it is largely blind to the purely quadrupolar orders that are even under time reversal. While recent progress proposed that quadrupolar excitations could become visible in inelastic neutron scattering~\cite{Zhentao_2025_PRL}, it requires the hybridization between one- and two-magnon channels. Moreover, this route still fundamentally relies on the spin-excitation spectrum. Here, we instead pursue an innovative phonon-based route, where bond phonons are not only a probe but also an active mechanism that enhance the stability of the spin-nematicity. Specifically, the emergence of spin nematic order imprints a distinctive signature onto the phonon spectrum through the spin-lattice coupling, providing a clear pathway for the detection of the hidden quadrupolar order.

We first demonstrate that the magnetoelastic couplings can substantially enhance and stabilize the spin nematic phase due to an induced biquadratic spin interaction. Then, we show that the emergence of quadrupolar order in the spin-nematic phase manifests itself in the phonon spectrum through two primary signatures, a pronounced phonon frequency splitting (an avoided crossing between phonon and magnon bands) and a strong dependence of the splitting position on the background spin-nematic order and the external magnetic field. Spectroscopic techniques such as Raman spectroscopy~\cite{Cui2023,Loudon01101964}, inelastic X-ray scattering~\cite{Burkel_1987, Krisch2007} are ideally suited to detect these changes, thereby translating the hidden spin order into an observable lattice response and offering a direct solution to the long-standing challenge of probing spin nematics. Our proposal is supported by both contemporary theoretical and experimental landscapes. Theoretically, the formation of the spin nematic phase via the condensation of two-magnon bound states is well established~\cite{Silberglitt_1970_PRB,Oguchi_1971_JPSP,Shannon_2013_PRB,Zhentao_2025_PRL}, with models highlighting the important role of anisotropy and frustration on the triangular lattice. Experimentally, the spin nematic order and two-magnon Bose-Einstein condensate have been reported in a growing set of candidate materials~\cite{Yoshimitsu_2019_PNAS, Kim2024, Bai2021,Sheng2025}, providing various platforms to test our proposal.

We start with a spin-1 XXZ model in a triangular lattice 
with the magnetoelastic interaction, where the lattice modulates the exchange couplings through vibrations. 
This framework will be used in two complementary ways, 
(i) integrating out the phonons to obtain an effective spin model 
and investigate how the spin–lattice coupling reshapes the spin states, 
and (ii) retaining phonons to compute the spectra with dynamical magnon-phonon couplings 
that provide the spectroscopic signatures of hidden spin nematicity~\cite{footnote}. 
Our model is written as
\begin{align}
\label{eq:Total_H}
    H=H_s + H_p + H_{me} ,
\end{align}
where $H_s$ and $H_p$ are the spin and phonon Hamiltonian, respectively. 
$H_{me}$ describes the magnetoelastic effect that characterizes the magnon-phonon interaction. 
The spin Hamiltonian includes a magnetic field $B$ and a large single-ion easy-axis anisotropy with
$D>0$ in addition to the XXZ anisotropy,
\begin{align}
\label{eq:H_spin_1}
    H_{s}=J\sum_{\braket{i,j}} (S_i^x S_j^x + S_i^y S_j^y+ \Delta S_i^z S_j^z)- D\sum_{i} (S_i^z)^2 -  B\sum_i S_i^z.
\end{align}
The exchange interaction is restricted between the nearest-neighbor sites 
with antiferromagnetic coupling $J>0$. $\Delta$ characterizes the XXZ anisotropy. 
In the model, we assume a large easy-axis anisotropy, 
which is often required for the emergence of on-site two-magnon bound state~\cite{Bai2021}. 
A moderate XXZ Ising anisotropy $\Delta>1$ is chosen. 
As for the phonon Hamiltonian $H_p$, in the bond-phonon model it takes the form
\begin{align}
    H_{p}= \sum_i \frac{|\bm{p}_i|^2}{2M} 
    + \frac{k_E}{2}\sum_{\braket{i,j}} \left(\bm{e}_{ij}\cdot \bm{u}_{ij}\right)^2,
\end{align}
where $M$ and $\bm{p}_i$ are the mass and momentum of the magnetic atom, respectively. 
${\bm{u}_{ij}\equiv \bm{u}_i-\bm{u}_j}$ is the in-plane lattice displacement 
and $\bm{e}_{ij}$ is the unit vector connecting sites $i$ and $j$. 
$k_E$ is the spring constant ${k_E=M\omega_0^2}$ with $\omega_0$ 
the intrinsic vibration frequency.

The magnetoelastic interaction can be obtained by expanding the exchange coupling 
with respect to the lattice distance $\bm{R}_{ij}$ around the equilibrium position 
$\bm{R}^0_{ij}$~\cite{Balents_2006_PRB, Gerrit_2019_PRB}, 
where ${J(\bm{R}_{ij}+u_{ij})\approx J(\bm{R}^0_{ij})
-\frac{\partial J}{\partial R} \bm{e}_{ij}\cdot \bm{u}_{ij}
\equiv J(1-\gamma\bm{e}_{ij}\cdot \bm{u}_{ij})}$. 
Therefore, the magnetoelastic interaction can be written as
\begin{align}
\label{eq:H_me}
    H_{me}=-J\gamma \sum_{\langle i,j \rangle} 
    \bm{e}_{ij} \cdot \bm{u}_{ij} (S_i^x S_j^x + S_i^y S_j^y + \Delta S^z_i S^z_j).
\end{align}
Here, the expansion coefficient $\gamma=(\partial_R J)/J$ 
is positive since the exchange usually decreases with increasing bond length. 

\begin{figure}[!t]
    \centering
    \includegraphics[width=1\linewidth]{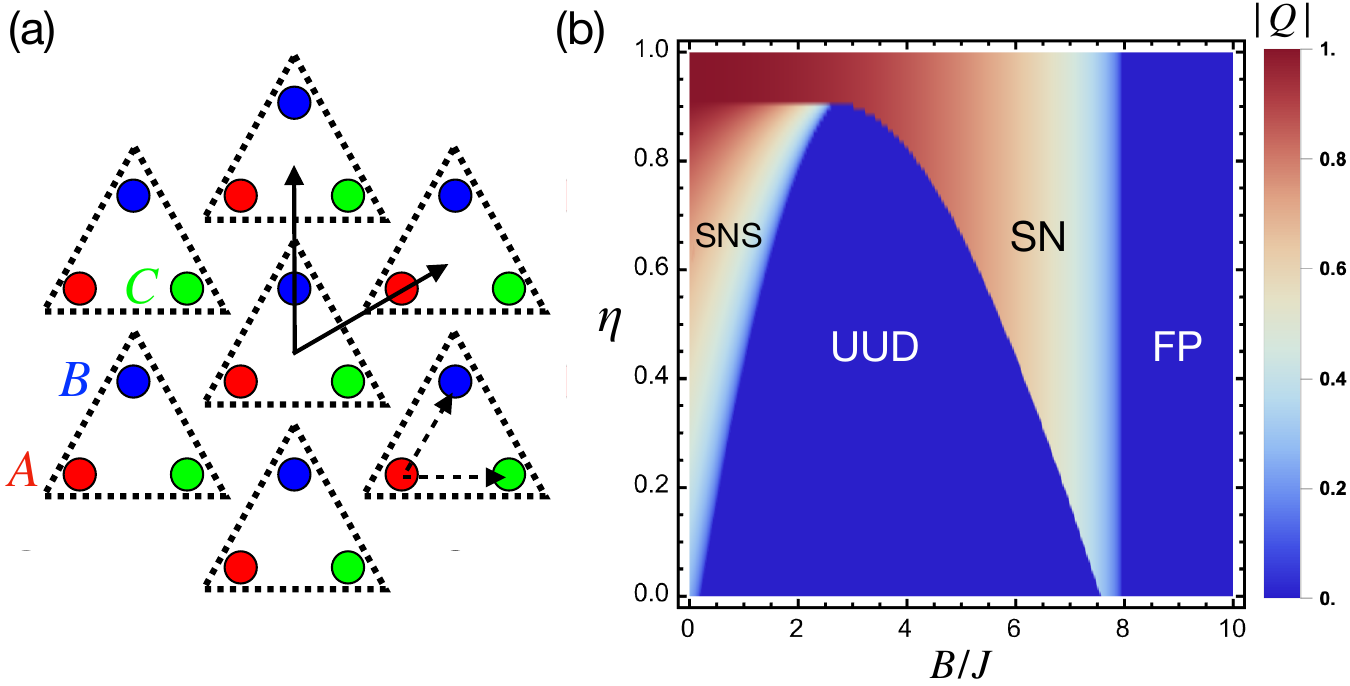}
    \caption{    
    (a) Triangular lattice structure with ordering wavevector at $K$ where a unit cell (dashed lines) 
    contains three sublattices as colored in red (A), blue (B) and green (C). 
    The black solid (dashed) arrows are the lattice vectors for the supercell (original triangular lattice).  
    (b) Phase diagram of $H_{s}^{\rm eff}$ in \eqref{eq:H_eff_SW} with the averaged quadrupolar order $|Q|\equiv  \sum_{\alpha\in\{A,B,C\}}{[\braket{Q^{x^2\!-\!y^2}_{\alpha}}^2+\braket{Q^{xy}_{\alpha}}^2]^{1/2}/3}$. Here and in the following, $D=4J$ and $\Delta=1.2$ are used in the calculation. 
    }
    \label{fig:lattice_phase}
\end{figure}

\noindent{\color{blue}\textit{Enhanced spin nematicity.}}---To explore the phonon effect
on the spin states, we integrate out the phonons $\bm{u}_{ij}$, 
that enters the Hamiltonian quadratically in $H_p$ and linearly in $H_{me}$. 
Consequently, the phonon field can be eliminated analytically via Gaussian 
integration~\cite{Balents_2006_PRB,SM}, resulting in a bilinear–biquadratic Hamiltonian
\begin{align}
\label{eq:H_eff}
    H_{s}^{\rm eff}=&J\sum_{\braket{i,j}} (S_i^x S_j^x + S_i^y S_j^y+ \Delta S_i^z S_j^z)- D\sum_{i} (S_i^z)^2 
        -  B\sum_i S_i^z\nonumber\\
    -&J_\eta\sum_{\braket{i,j}} (S_i^x S_j^x + S_i^y S_j^y+ \Delta S_i^z S_j^z)^2,
\end{align}
where a biquadratic term emerges with the strength $J_\eta \equiv J\eta$. Here, the dimensionless magnetoelastic coupling coefficient is defined as $\eta=J\gamma^2/k_E$.
While such biquadratic term is known to be essential for the quadrupolar spin order~\cite{Hsieh_1969_JAP,Penc_2006_PRL, Moreno_2014_PRB}, pinning down its precise microscopic origin in real materials remains a challenge. 
Various mechanisms have been suggested, ranging from the twisted ring exchange~\cite{Tanaka_2018_JPSP}, 
higher-order perturbation from spin-charge coupling~\cite{Hayami_2017_PRB,Akagi_2012_PRL}, 
virtual crystal field fluctuations~\cite{Rau_2016_PRB} and multipolar interactions~\cite{Santini_2009_RMP}. 
Here, the biquadratic term in Eq.~\eqref{eq:H_eff} arises naturally from magnetoelastic interactions~\cite{Balents_2006_PRB,Tchernyshyov_2002_PRL}.


To clarify how the quadrupolar order emerges from the effective model in Eq.~\eqref{eq:H_eff}, 
we further project out the high-energy $\ket{S^z=0}$ state to establish a low-energy effective model
in the manifold of $\{\ket{\uparrow}\equiv \ket{S^z=+1},\ket{\downarrow}\equiv \ket{S^z=-1} \}$,
that is demanded by the easy-axis single-ion anisotropy. Within this two-fold local Hilbert space~\cite{Zhentao_2025_PRL},
the pseudospin operators $\bm{\tau}$ 
are introduced with $\tau^z=PS^zP/2,\ \tau^{\pm}=P(S^{\pm})^2P/2$, 
where $S^{\pm}=S^x\pm \mathrm{i}S^y$, $\tau^{\pm}=\tau^x\pm \mathrm{i}\tau^y$, 
and $P=\ket{\uparrow}\bra{\uparrow} + \ket{\downarrow}\bra{\downarrow}$ 
is the projection operator for the pseudospin-1/2 space. It
is obvious that $PS^{x,y}P=0$. 
Therefore, if we directly project the Hamiltonian~\eqref{eq:H_eff} for each site, 
the pairwise exchange term $S_i^x S_j^x + S_i^y S_j^y$ will not survive. 
However, such one-magnon process can be treated as perturbation 
and after a Schrieffer–Wolff transformation for projecting into the low-energy space, 
it becomes biquadratic~\cite{SM}, which has the same form as the last term in Hamiltonian~\eqref{eq:H_eff}. 
The final effective Hamiltonian then takes the form~\cite{SM}
\begin{align}
\label{eq:H_eff_SW}
    H_{s}^{\rm eff}=\sum_{\langle i,j\rangle } 
    \left[J_{xy}(\tau^x_i\tau^x_j + \tau^y_i\tau^y_j) + J_{z}\tau^z_i \tau^z_j \right] - 2 B\sum_i \tau_i^z,
\end{align}
where we find that the two exchange anisotropies are now renormalized 
by the magnetoelastic interaction as
\begin{subequations}
\label{eq:Delta}
\begin{align}
    J_{xy} &=-\frac{J^2}{D}-2J_\eta + \frac{(J_\eta\Delta)^2}{D}\label{eq:Jxy},\\
    J_z&=\frac{J^2}{D}+ 4J\Delta -2J_\eta + \frac{(J_\eta \Delta)^2}{D}.
\end{align}
\end{subequations}
The quadrupolar nature of the Hamiltonian~\eqref{eq:H_eff_SW} 
becomes explicit, as the in-plane terms can be recast as
\begin{align}
    J_{xy}(\tau^x_i\tau^x_j + \tau^y_i\tau^y_j)
    =\frac{J_{xy}}{4} P(Q^{x^2\!-\!y^2}_{i}Q^{x^2\!-\!y^2}_{j} + Q^{xy}_{i}Q^{xy}_{j})P
    \label{eq:b_and_Q}.
\end{align}
Here, the on-site spin quadrupolar orders are defined as 
$Q^{x^2\!-\!y^2}_{i} =(S_i^x)^2-(S_i^y)^2$ and 
$Q^{xy}_{i}= S_i^xS_i^y+S_i^yS_i^x$, which satisfy the relations $PQ_i^{x^2\!-\!y^2}P=2\tau_x$ and $PQ_i^{xy}P=2\tau_y$. 
A quadrupolar order $\braket{Q}$ in the original spin-1 model~\eqref{eq:H_spin_1} 
is then translated to the dipole order $\braket{\bm{\tau}}$ in the pseudospin-1/2 model~\eqref{eq:H_eff_SW}. 
The quadrupolar order originates from the two-magnon processes ($S^+ S^+$ and $S^- S^-$) 
that connect the local ground states $\ket{S_z=+1}$ and $\ket{S_z=-1}$. 
From Eq.~\eqref{eq:Delta}, we find that $J_{xy}$ and $J_{z}$ 
share the same critical value $\eta_c=D/J\Delta^2$, 
below which ($0<\eta<\eta_c$), 
increasing $\eta$ results in an increase in $|J_{xy}|$ and a decrease in $|J_{z}|$. Specifically, a finite $\eta$ can greatly enhance $|J_{xy}|$ while only moderately suppressing $|J_z|$. This asymmetry arises because the bare terms (at $\eta=0$) are of second and first order in the pairwise exchange $J$ for $J_{xy}$ and $J_{z}$, respectively. Consequently, a physically accessible spin-lattice coupling ($\eta<1<\eta_c$) is expected to enhance the stability of the quadrupolar order by boosting the ratio $|J_{xy}/J_{z}|$. To justify this, we solve for the full phase diagram that is depicted in Fig.~\ref{fig:lattice_phase}. We use the site-factorized $SU(2)$ coherent state representation to minimize the energy of the effective Hamiltonian~\eqref{eq:H_eff_SW}. Such a semiclassical method has 
capture all the phases calculated by density matrix renormalization group and infinite projected entangled pair states, 
with only a slight shift of the critical fields~\cite{Zhentao_2025_PRL,Yamamoto_2014_PRL}.

The phase diagram is shown in Fig.~\ref{fig:lattice_phase}, where four distinct phases,
spin-nematic-supersolid (SNS), up-up-down (UUD), spin-nematic (SN) and fully polarized (FP) phases,
are obtained. For the SN and FP phases, the ordering wavevector is at $\Gamma$. For both SNS and UUD phases,
the ordering wavevector is at $K$ and the magnetic unit cell has three sublattices (see Fig.~\ref{fig:lattice_phase}). 
Up to a global rotation in the $xy$ plane (Hamiltonian~\eqref{eq:H_eff_SW} has U(1) symmetry), 
the directions of the effective spins $\bm{\tau}_\alpha$ for different phases are found to be
\begin{small}
\begin{subequations}
\begin{align}
    \text{SNS}:&\hat{\bm{\tau}}_A=\hat{\bm{\tau}}_B=(\sin\theta_1,0,\cos\theta_1),\ \hat{\bm{\tau}}_C=(\sin\theta_2,0,\cos\theta_2) , \\
    \text{UUD}:&\hat{\bm{\tau}}_A=\hat{\bm{\tau}}_B=(0,0,1),\  \hat{\bm{\tau}}_C=(0,0,-1) , \\
    \text{SN}:&\hat{\bm{\tau}}_A=\hat{\bm{\tau}}_B=\hat{\bm{\tau}}_C=(\sin\phi,0,\cos\phi), \label{eqsn} \\
    \text{FP}:&\hat{\bm{\tau}}_A=\hat{\bm{\tau}}_B=\hat{\bm{\tau}}_C=(0,0,1). 
\end{align}
\end{subequations}
\end{small}
Here, the angles $\theta_1,\theta_2$ and $\phi$ are determined by minimizing the ground state energy~\cite{SM}. 
In contrast to the SN phase, the SNS phase breaks the lattice translation symmetry of the original triangular 
lattice in a characteristic $\sqrt{3}\times\sqrt{3}$ pattern~\cite{Heidarian_2005_PRL}.

In Fig.~\ref{fig:lattice_phase}, the regions for the SNS and SN phases are quite narrow 
at $\eta=0$~\cite{Zhentao_2025_PRL}, since $J_{xy}(\eta=0)=J^2/D$ only has the high-order contribution 
from the bilinear pair-wise exchange term of Eq.~\eqref{eq:H_eff}. 
In the parameter choice of Fig.~\ref{fig:lattice_phase},
the three critical fields at $\eta=0$ are $B_{c,1}/J \approx 0.2 $, $B_{c,2}/J= 7.52 $, $B_{c,3}/J=7.95$. 
In the low-field regime ($B<B_{c,1}$), the system stabilizes in the SNS state. 
Upon entering the intermediate field range ($B_{c,1}<B<B_{c,2}$), 
it exhibits UUD spin order. As $B$ increases further ($B_{c,2}<B<B_{c,3}$), 
the system transitions into the SN phase. Finally, above the saturation field (${B > B_{c,3}}$), 
the spins become fully polarized (FP phase). 
When $\eta$ increases, $B_{c,1}$ shifts to higher values and $B_{c,2}$ decreases, 
whereas $B_{c,3}$ remains essentially unchanged (independent of $\eta$). 
Consequently, a finite $\eta$ {\sl substantially enlarges} the phase regions of both the SNS and SN states, 
allowing them to persist over a much broader range of magnetic fields as shown in Fig.~\ref{fig:lattice_phase}. 
For a sufficiently large $\eta  \gtrsim 0.9$, 
the SNS phase is no longer stable and will be overtaken by the SN phase. 
An order-of-magnitude estimation for $\eta$ is left in supplementary material (SM)~\cite{SM}. 
In both dipole-ordered phases (UUD and FP), the quadrupolar order parameter $|Q|$ vanishes identically, 
whereas $|Q|$ is finite only in the SNS and SN phases. Current experimental techniques, however, 
are primarily sensitive to the spin dipolar order $\braket{\bm{S}}$, and cannot directly detect the quadrupolar order $|Q|$. Therefore, we next examine how the quadrupolar order of the SNS and SN phases 
affect the phonon spectra, leading to the practical route for phononic detection of hidden spin nematicity.

\noindent{\color{blue}\textit{Spin-nematic phase.}}---To investigate the phonon spectrum 
with finite quadrupolar order in the SN phase, we start from the total Hamiltonian $H=H_s+H_p +H_{me}$ 
without integrating out the phonon fields, 
which can be written in the Nambu basis $\psi =(b_{1,\bm{k}},b_{2,\bm{k}},b_{1,-\bm{k}}^\dagger,b^\dagger_{2,-\bm{k}},a_{\bm{k}}, a^\dagger_{-\bm{k}})^T$ as $H=\frac{1}{2}\sum_{\bm{k}}\psi^\dagger  \mathcal{H}^{SN} \psi$. 
The creation operators for phonon and magnon are denoted by $b^\dag$ and $a^\dag$, respectively. 
The kernel $\mathcal{H}^{SN}$ takes a block matrix form~\cite{SM}
\begin{align}\label{eq:H_tot_SN}
    \mathcal{H}^{SN}=
    \begin{pmatrix}
        \mathcal{H}_p(\bm{k}) & \Pi^\dagger(\bm{k}) \\
        \Pi(\bm{k}) & \mathcal{H}_s(\bm{k})
    \end{pmatrix},
\end{align}
where $\mathcal{H}_p$ and $\mathcal{H}_{s}$ are the kernels for free phonon and magnon, respectively. $\Pi(\bm{k})$ originates from $H_{me}$~\cite{SM}. 
By transforming the Hamiltonian $\mathcal{H}^{SN}_{p\text{-}h}=(\Lambda^{-1})^\dag \mathcal{H}^{SN}\Lambda^{-1} $ to the particle-hole symmetric basis $\Psi=\Lambda \psi=(b_{1,\bm{k}},b_{2,\bm{k}},a_{\bm{k}},b_{1,-\bm{k}}^\dagger,b^\dagger_{2,-\bm{k}}, a^\dagger_{-\bm{k}})^T$, the diagonalization of the Hamiltonian $Q^\dag \mathcal{H}^{SN}_{p\text{-}h} Q=\varepsilon $ can be mapped to a generalized eigenvalue problem $\mathcal{H}^{SN}_{p\text{-}h} Q=gQ \bar{\varepsilon}$ with the diagonal energy matrix $\bar{\epsilon}\equiv g\varepsilon$ and metric $g=\sigma_z\otimes I_{3\times3}$. 
Here the Bogoliubov transformation $Q$ must satisfy the condition $Q^\dagger g Q=Qg Q^\dagger =g$ 
to maintain the correct bosonic commutation relation for magnon and phonon operators. 

By writing the Hamiltonian in a quadratic form~\eqref{eq:H_tot_SN}, 
we have required the dangling (linear) terms in $H_s$ to vanish identically, 
whose presence indicates an unstable magnetic ground state. 
This condition yields two solutions for $\phi$ in Eq.~\eqref{eqsn}. 
The trivial solution $\phi= 0$ corresponds to the FP phase. 
The nontrivial solution $\phi= \arccos[2 B D /3J^2(2+\frac{4\Delta D}{J})]$ 
with lower energy corresponds to the SN phase. 
The spin configuration defined by this nontrivial solution exactly 
matches with the one obtained by minimizing energy in the SN phase at $\eta=0$. 
The existence of this non-trivial solution imposes an upper bound of the magnetic field 
$B_{max}=3J^2(2+\frac{4\Delta D}{J})/2 D $ for the SN phase, 
which is found to coincide with the upper boundary $B_{c,3}=B_{max}$ 
in Fig.~\ref{fig:lattice_phase}.

\begin{figure}[!t]
    \centering
    \includegraphics[width=1\linewidth]{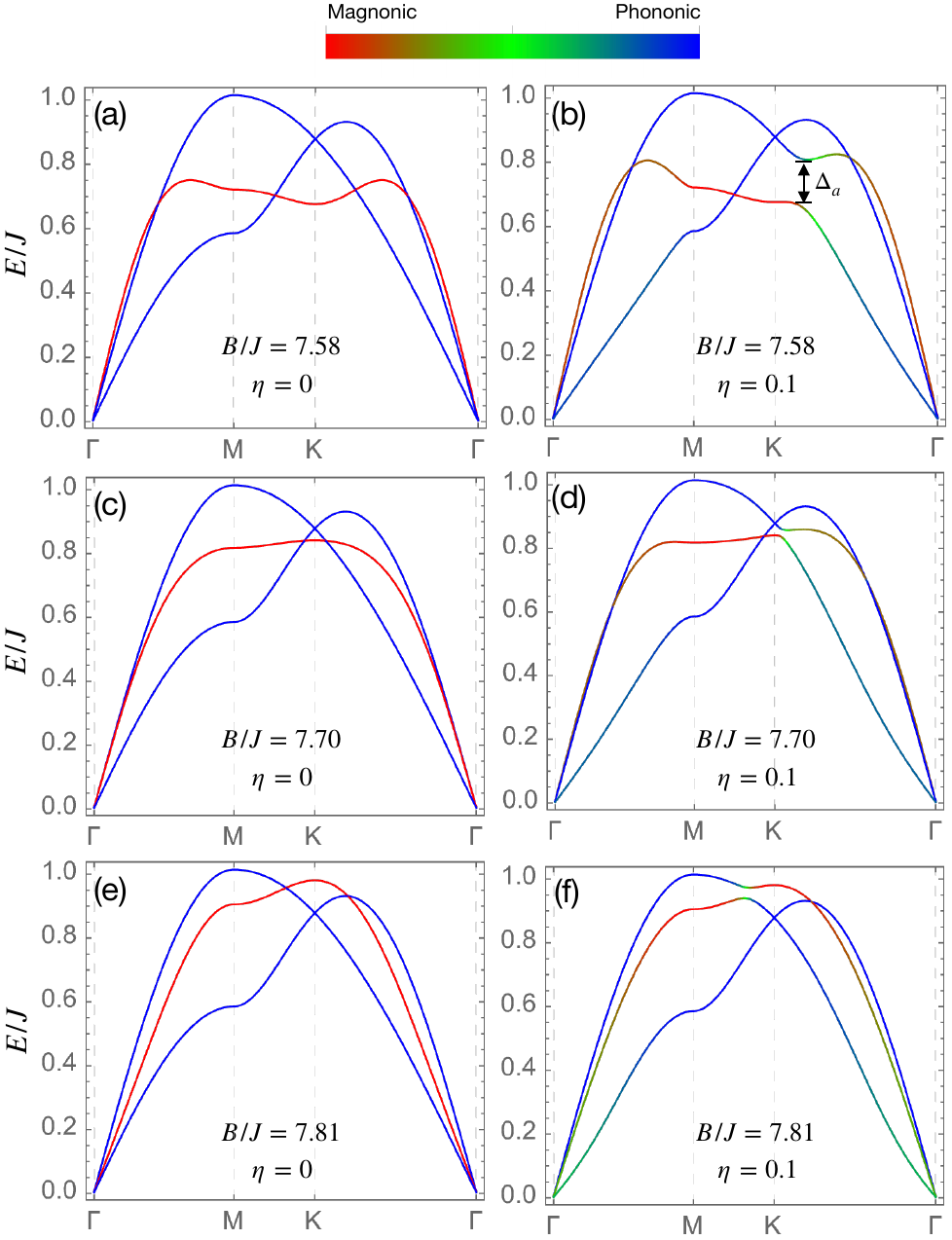}
    \caption{Phonon–magnon band spectra in SN phase under different magnetic fields in the absence [(a), (c), and (e)] and presence [(b), (d), and (f)] of magnon–phonon coupling. The color represents the normalized relative weight of the phonon-like and magnon-like components of each band.  The intrinsic frequency $ \hbar \omega_0/ J=0.41$. }
    \label{fig:SN-bands}
\end{figure}

In the SN phase with a monoatomic 2D unit cell, 
the spectrum comprises one magnon band and two low-energy acoustic phonon bands (no optical modes). 
The calculated energy spectra of Hamiltonian~\eqref{eq:H_tot_SN} are presented in Fig.~\ref{fig:SN-bands} 
with a low intrinsic frequency for the acoustic phonon. 
In the first column ($\eta=0$) of Fig.~\ref{fig:SN-bands}, 
the magnon dispersion is situated between the two phonon branches. 
The gapless feature of magnon bands (red line) is consistent with the spontaneous $U(1)$ symmetry breaking due to
the finite in-plane dipole order $\braket{\tau}$.

\begin{figure}[!t]
    \centering
    \includegraphics[width=0.8\linewidth]{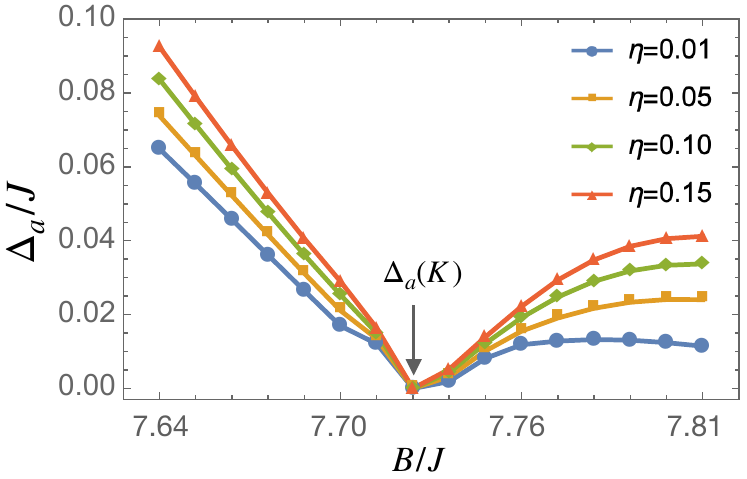}
    \caption{Magnitude of the avoided crossing gap $\Delta_{a}$ as a function of magnetic field in the SN phase. The black arrow indicates the critical magnetic field $B/J\approx 7.72$ where $\Delta_{a}$ vanishes at the $K$ point.}
    \label{fig:gap}
\end{figure}

In the presence of finite magnon-phonon coupling ($\eta\neq 0$), 
an avoided band gap opens at the energy crossings between the magnon and phonon bands as shown in the second column of Fig.~\ref{fig:SN-bands}. 
Several key features are worth noting. 
First, a finite $\eta$ does not lift the degeneracy of phonon bands at the $\Gamma$ and $K$ points, as the magnetoelastic interaction $H_{me}$ vanishes identically at these high-symmetry points. 
Second, although there are multiple crossings between the magnon and phonon bands, only one avoided crossing emerges prominently. 
Third, the position of this gap is very sensitive to the magnetic field. 
As we gradually increase the magnetic field, the position of the avoided crossing sweeps through the $K$ point. 
Since the gap is enforced to close exactly at $K$ (due to vanishing $H_{me}$), we expect the magnitude of this avoided crossing, denoted as $\Delta_{a}$, to exhibit a non-monotonic dependence on the magnetic field. 
In Fig.~\ref{fig:gap}, we plot $\Delta_{a}$ as a function of magnetic field for various coupling strengths $\eta$. 
It is clear that $\Delta_{a}$ increases with increasing $\eta$. 
Most importantly, there exists a critical magnetic field where the crossing position can be tuned to be right at the $K$ point, 
yielding a vanishing gap $\Delta_{a}(K)=0$ (Fig.~\ref{fig:gap}). 
Tuning the magnetic field away from this critical point reopens an observable gap. Consequently, $\Delta_{a}$ demonstrates a non-monotonic dependence on $B$, 
as anticipated. All the non-trivial features of the phonon bands discussed 
above serve as a distinct signature for probing the SN phase.

\noindent{\color{blue}\textit{Spin-nematic-supersolid phase.}}---In the SNS phase, 
the unit cell contains three sublattices (Fig.~\ref{fig:lattice_phase}). 
Thus, we enlarge the unit cell for phonons to include the magnon-phonon coupling. 
We obtain six phonon bands in total and four of them are the fictitious optical modes 
due to the band folding, which can be projected out by the projector $P^{ac}$ 
constructed specifically for the acoustic modes~\cite{SM}.

\begin{figure*}[!t]
    \centering
    \includegraphics[width=0.8\linewidth]{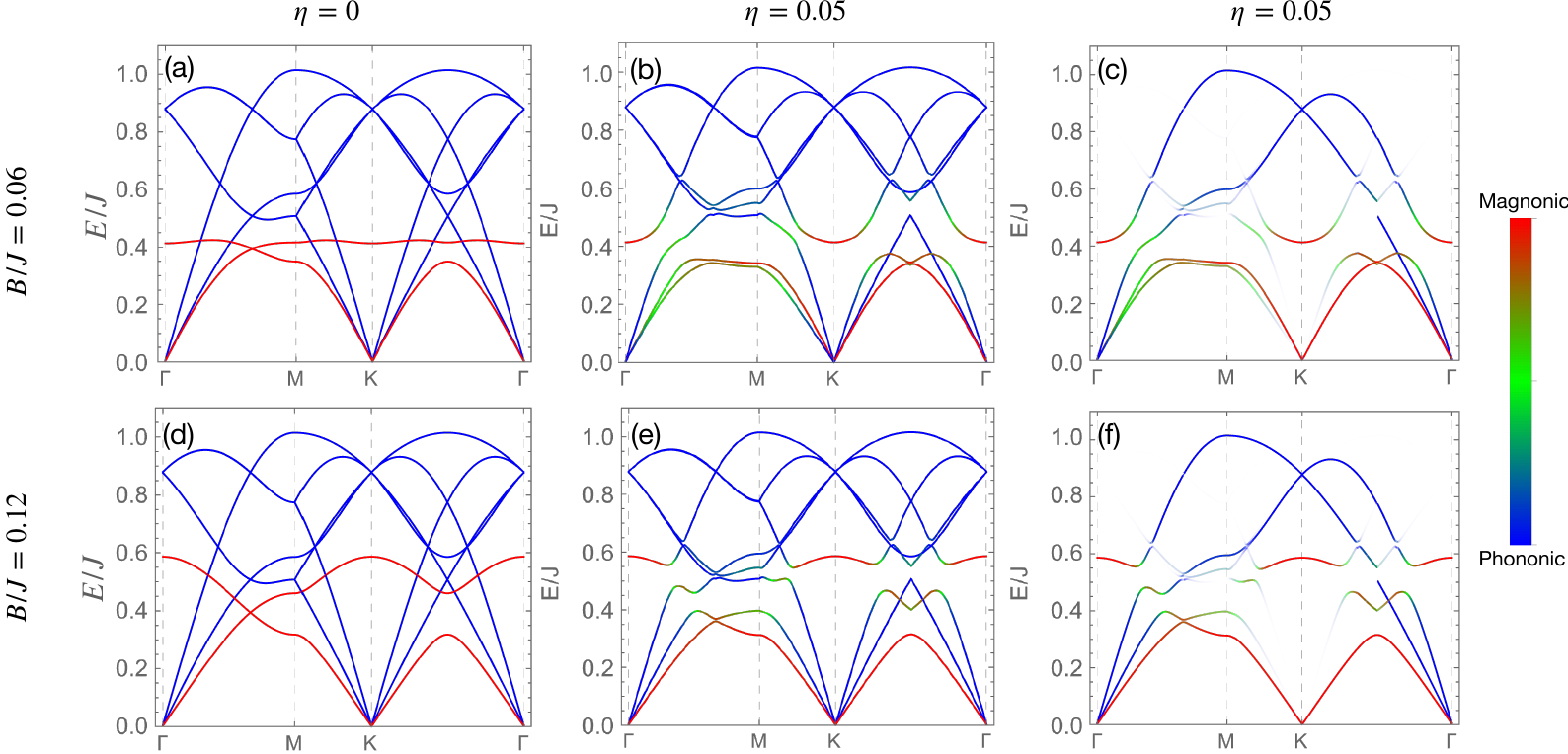}
    \caption{Phonon–magnon band spectra in SNS phase with magnetic field $B/J=0.06$ [first row, (a-c)] and $B/J=0.12$ [second row, (d-f)]. In the first column [(a) and (d)], the magnon-phonon coupling is turned off ($\eta=0$). In the second column [(b) and (e)], the magnon-phonon coupling is turned on ($\eta\neq 0$). (c) and (f) are the same as (b) and (e) but with the fictitious optical modes projected out. The intrinsic frequency $\omega_0$ takes the same value as in Fig.~\ref{fig:SN-bands}.}
    \label{fig:SNS-bands}
\end{figure*}

Following the same procedure as in the SN phase, we find that the vanishing of dangling (linear) 
terms in $H_s$ yields two transcendental equations for the angle $\theta_1$ and $\theta_2$~\cite{SM}. 
The trivial solution $\theta_1=0,\theta_2=\pi$ corresponds to the UUD phase while the nontrivial solution corresponds to the SNS phase, which exactly matches the one obtained from minimizing the energy in the SNS phase. Again, the existence of such nontrivial solution imposes an upper bound $B_{max}$ of the magnetic field 
for the SNS phase, which coincides with the threshold $B_{max}=B_{c,1}$ of Fig.~\ref{fig:lattice_phase}(b). 
After obtaining a stable magnetic ground state, we plot the phonon and magnon band spectra 
for different magnetic fields in Fig.~\ref{fig:SNS-bands}.

In stark contrast to the SN phase where the gap opens at a single crossing point, 
the SNS phase exhibits multiple avoided band gaps across the spectrum. 
This complexity arises because $H_{s}$ and $H_{me}$ highly depend on the underlying spin order and the magnetoelastic interaction in the SNS phase allows richer couplings. 
The color-coded spectral weights in Figs.~\ref{fig:SNS-bands}(b) and (e) reveal strong hybridization 
at all the crossings between magnons and phonons. With increasing magnetic field, 
the avoided band gaps shift to higher energy. The hybridization between magnons 
and fictitious optical phonons occurs around the $M$ and $K$ points in the middle of the energy spectra. 
The two low-energy phonon bands at the $K$ point originate from band folding associated 
with the enlarged unit cell adopted for the phonon. In Figs.~\ref{fig:SNS-bands}(c) and (f), 
we show the band spectra where such fictitious optical bands are removed by the projector $P^{ac}$. 
Importantly, the magnon band can display splittings even in regions devoid of 
genuine acoustic phonon modes [Figs.~\ref{fig:SNS-bands}(c) and (f)]. 
Physically, these gaps originate from the interaction between magnons 
and the folded phonon branches (Umklapp scattering). 
The degeneracies of phonon bands at $K$ and $\Gamma$ points 
cannot be lifted by a finite $\eta$.

In Fig.~\ref{fig:SNS-bands}, only two magnon bands are shown,
 while a high-energy flat band (around $E/J\sim 14.5$), 
 which is insensitive to the magnetic field, is not explicitly displayed. 
 The emergence of this high-energy flat band can be attributed to the geometrical 
 frustration of the triangular lattice in the $J_{z}\gg J_{xy}$ Ising limit~\cite{Wannier_1950_PR}. 
 We find that this flat band is completely decoupled from the lattice degrees of freedom, even when the phonon energies are artificially increased to force energy crossings for the magnon flat band, a finite ($\eta\neq 0$) fails to open any spectral gap.

\noindent{\color{blue}\textit{Discussion.}}---We conclude that the spin–lattice coupling plays two key roles in our work. First, it substantially enhances the stability of the SN and SNS phases. 
Second, it induces avoided crossings in the phonon spectrum that depend on 
the underlying spin ground state, enabling a practical phononic probe of the otherwise 
hidden quadrupolar order. 
The implications of successfully detecting spin nematicity via phonons are profound 
and open several new research avenues.
 Our methodology could be generalized to probe other elusive forms of hidden orders,
  such as those found in spin liquids~\cite{Wen_2002_PRB,Zhou_RMP_2017} or other multipolar states~\cite{Santini_2009_RMP,Patri2019}, 
  across different lattice geometries. 
  Furthermore, studying the interplay between phonons and the dynamical quadrupolar structure factor 
  could yield fresh insights into the low-energy excitations of the multipolar order~\cite{Liu_2018_PRB,hart2025phonondriven}. 
  
  Beyond the solid-state systems, these concepts could be tested in tunable quantum simulator platforms, 
  such as spin-1 Bose-Einstein condensates~\cite{Sheng2025}, 
  where spin-spin correlations can be engineered and observed with high precision. 
  Thus, leveraging the bond phonon model not only solves an immediate detection problem 
  but also establishes a versatile new paradigm for exploring the rich landscape of complex 
  quantum magnetism. Recently, we became aware of related work~\cite{sutcliffe2025} 
  that employs on-site (Einstein) phonons to probe spin-orbit-entangled multipolar order 
  through the lifting of the $E_g$-doublet degeneracy; 
  by contrast, our bond-phonon mechanism can both enhance 
  and diagnose a purely-spin quadrupolar order via robust magnon-phonon hybridization (avoided crossings),
   without lifting the phonon degeneracies.

\noindent\emph{Acknowledgments.}---We thank Jiahao Yang for the fruitful discussions. 
This work is supported by NSFC No.92565110, No.12574061, 
and by the Ministry of Science and Technology of China with Grants No.~2021YFA1400300.

\bibliography{ref}

\clearpage
\onecolumngrid

\let\addcontentslineorig\addcontentsline
\renewcommand{\addcontentsline}[3]{%
  \def\tempa{#2}%
  \def\sec{section}%
  \def\subsec{subsection}%
  \def\subsubsec{subsubsection}%
  \ifx\tempa\sec
    \addcontentslineorig{stoc}{#2}{#3}%
  \else\ifx\tempa\subsec
    \addcontentslineorig{stoc}{#2}{#3}%
  \else\ifx\tempa\subsubsec
    \addcontentslineorig{stoc}{#2}{#3}%
  \else
  \fi\fi\fi
}

\setcounter{section}{0} 
\setcounter{equation}{0}
\renewcommand{\theequation}{S\arabic{equation}}
\renewcommand\thefigure{S\arabic{figure}}
\setcounter{figure}{0}
\begin{center}
    \textbf{}
    \textbf{}
    \textbf{\large Supplementary material for ``Phononic enhancement and detection of hidden spin-nematicity and dynamics in quantum magnets''}\\[10pt]
    Junyu Tang$ ^1$, Hong-Hao Song$ ^{1}$, Gang v.~Chen$ ^{1,2}$ \\[10pt]
    \textit{$^1$International Center for Quantum Materials, School of Physics, Peking University, Beijing 100871, China}\\[10pt]
    \textit{$^2$Collaborative Innovation Center of Quantum Matter, 100871, Beijing, China}
\end{center}
\onecolumngrid
\supplementarytableofcontents

\section{Schrieffer–Wolff transformation}
The spin-1 XXZ Hamiltonian [Eq.(2) of main text] can be divided into as $H=H_0 + V$, where the perturbation $V=J\sum_{\braket{i,j}} (S_i^x S_j^x + S_i^y S_j^y)=J\sum_{\langle i,j \rangle} (S_i^+S_j^-+S_i^-S_j^+)/2$ corresponds to the spin flipping and thus belongs the high energy part ($D\gg J, g_c\mu_B B$). The Schrieffer–Wolff transformation seeks for a transformation $H'=e^{S} He^{-S}=H_0 + \frac12 [S,V] + O(V^3)$ that projects into the effective low-energy space. The generator $S$ should satisfy the condition $V+[S,H_0]=0$. Most importantly, $H_0$ is diagonal while the perturbation is purely off-diagonal ($V_{nn}=0$) in the low-energy space. Therefore, we have the following solution for the generator $S$:
\begin{align}\label{seq:SV}
    \frac{1}{2}[S,V]_{mn} =\frac{1}{2}\sum_k V_{mk}V_{kn} [\frac{1}{E_m-E_k}-\frac{1}{E_k-E_n} ],
\end{align}
where $O_{nm}\equiv \braket{n|\hat{O}|m}$ and $E_n$ is the energy of state $\ket{n}$. After plugging the expression of $V$ into above equation, we find the following four terms for $V_{mk}V_{kn}$
\begin{align}\label{seq:VV}
    V_{mk}V_{kn} &= \frac{J^2}{4}\langle m|S^+_{i}S_j^-|k\rangle \langle k|S^+_{i}S_j^-|n\rangle + + \frac{J^2}{4}\langle m|S^-_{i}S_j^+|k\rangle \langle k|S^-_{i}S_j^+|n\rangle\nonumber\\
    &+ \frac{J^2}{4}\langle m|S^+_{i}S_j^-|k\rangle \langle k|S^-_{i}S_j^+|n\rangle + \frac{J^2}{4}\langle m|S^-_{i}S_j^+|k\rangle \langle k|S^+_{i}S_j^-|n\rangle 
\end{align}
If we denote the state at site $i,j$ as $$\ket{S_i^z,S^z_j}\equiv \ket{S_i^z} \ket{S_j^z},\  S_z=\pm 1 \to \{1,\bar{1}\}$$, then it's easy to see that the first term only has a contribution from $|m\rangle =| 1,\bar 1\rangle ,|k\rangle=|0,0\rangle  , |n\rangle =|\bar 1,1\rangle $. Similarly, the second term only has a contribution from $|m\rangle =|\bar{1}, 1\rangle ,|k\rangle=|0,0\rangle  , |n\rangle =|1,\bar 1\rangle$. Combining with Eq.~\eqref{seq:SV}, the first two terms of Eq.~\eqref{seq:VV} yield a Hamiltonian containing $-\frac{J^2}{2D} |1,\bar 1\rangle \langle \bar 1,1|$ and $-\frac{J^2}{2D} |\bar 1, 1\rangle \langle 1,\bar 1| $, which corresponds to a biquadratic Hamiltonian 
\begin{align}
    -\frac{J^2}{8D} \sum_{\langle i,j \rangle} (S^+_iS^+_i S_j^- S_j^- + S^-_iS^-_i S_j^+ S_j^+)
\end{align}

The last two terms of Eq.~\eqref{seq:VV} need a careful examination, where we find that a nonvanishing contribution requires $\ket{m}=\ket{n}$. However, among the four possible states of $\ket{m}$ and $\ket{n}$ for $\langle m|S^+_{i}S_j^-|k\rangle \langle k|S^-_{i}S_j^+|n\rangle$, only $|n\rangle =|1,\bar 1\rangle,|m\rangle =|1,\bar 1\rangle$ are relevant to us since they reside in the desired low energy space, which is connected through an intermediate high-energy virtual state $\ket{k}=\ket{0,0}$. Similarly, for $ \langle m|S^-_{i}S_j^+|k\rangle \langle k|S^+_{i}S_j^-|n\rangle$, we only has the contribution from $|n\rangle =|\bar 1, 1\rangle ,|k\rangle=|0,0\rangle ,|m\rangle =|\bar 1, 1\rangle$. Combining with Eq.~\eqref{seq:SV}, the last two terms of Eq.~\eqref{seq:VV} yield a Hamiltonian containing $-\frac{J^2}{2D} |1,\bar 1\rangle \langle 1,\bar 1|$ and $-\frac{J^2}{2D} |\bar 1, 1\rangle \langle \bar 1, 1|$, which corresponds to the operator $\frac{J^2}{4D} S^z_i S^z_j $ plus some constant coefficient proportional to the identity operator that can be neglected.  

To conclude, after performing the Schrieffer–Wolff transformation, The bilinear term $J\sum_{\langle i,j \rangle} (S_i^+S_j^-+S_i^-S_j^+)/2$ in $H_s$ contributes to a Hamiltonian in the form of 
\begin{align}
    -\frac{J^2}{8D} \sum_{\langle i,j \rangle} (S^+_iS^+_i S_j^- S_j^- + S^-_iS^-_i S_j^+ S_j^+)+\frac{J^2}{4D} S^z_i S^z_j 
\end{align}
In the next section, we will see how phonon affects the coefficient of the bilinear term and meanwhile brings the additional biquadratic term.

\section{Effective spin Hamiltonian}
The phonon field $\bm{u}_{ij}$ enters the Hamiltonian quadratically in $H_p$ and linearly in $H_{me}$. Additionally, each $\bm{u}_{ij}$ couples
only to a single nearest-neighbor pair of spins, and can be treated as an independent variable\cite{Balents_2006_PRB}. Consequently, the phonon field can be eliminated analytically by a Gaussian integration:
\begin{align}
    e^{-\beta H_s^{\rm eff}}=\prod_{\braket{i,j}}\int \mathcal{D}\bm{u}_{ij} e^{-\beta(H_s + H_p[\bm{u}_{ij}] + H_{me}[\bm{u}_{ij}])},
\end{align}
where $\beta = 1/k_B T$. We note that completing the square in the exponent is physically equivalent to replacing the $\bm{u}{ij}$ field with the equilibrium value that minimizes the total energy $H_p[\bm{u}_{ij}] + H_{me}[\bm{u}_{ij}]$. Thus, up to a constant term, one obtains
\begin{align}
        H_{s}^{\rm eff}=H_s
    -J\eta\sum_{\braket{i,j}} (S_i^x S_j^x + S_i^y S_j^y+ \Delta S_i^z S_j^z)^2,
\end{align}
which is Eq.(5) in the main text. Notably, the biquadratic term emerges in the effective Hamiltonian after integrating out the phonon field. Expanding the square of this biquadratic term, we find
\begin{align}
    -J\eta\sum_{\braket{i,j}} (S_i^x S_j^x + S_i^y S_j^y+ \Delta S_i^z S_j^z)^2 &= - \frac{J\eta}{4}\sum_{\langle i,j\rangle} (S_i^+S_j^-+S_i^-S_j^+)^2 + 2\Delta S_i^z S_j^z (S_i^+S_j^-+S_i^-S_j^+) + 2\Delta  (S_i^+S_j^-+S_i^-S_j^+) S_i^z S_j^z\nonumber\\
    &=- \frac{J\eta}{4}\sum_{\langle i,j\rangle} (S_i^+S_i^+S_j^-S_j^-+S_i^-S_i^-S_j^+S_j^+ + S_i^+S_i^-S_j^-S_j^++S_i^-S_i^+S_j^+S_j^-)\nonumber\\
    &-J\eta \Delta \sum_{\langle i,j\rangle} S_i^z S_j^z \frac{(S_i^+S_j^-+S_i^-S_j^+)}{2} -J\eta\Delta\sum_{\langle i,j\rangle}  \frac{(S_i^+S_j^-+S_i^-S_j^+)}{2}S_i^z S_j^z 
\end{align}
where we have ignore the term $\Delta^2 (S^z_i)^2 (S^z_j)^2 $, which corresponds to a constant term in the low-energy space. The effective spin Hamiltonian now becomes
\begin{align}
    H_{s}^{\rm eff}=&J\sum_{\langle i,j \rangle} \left[\frac{(1-\eta\Delta S^z_i S^z_j)(S_i^+S_j^-+S_i^-S_j^+)-(S_i^+S_j^-+S_i^-S_j^+)\eta\Delta S^z_i S^z_j}{2} \right]\nonumber\\
    &- \frac{J\eta}{4}\sum_{\langle i,j\rangle} (S_i^+S_i^+S_j^-S_j^-+S_i^-S_i^-S_j^+S_j^+ + S_i^+S_i^-S_j^-S_j^++S_i^-S_i^+S_j^+S_j^-)\nonumber\\
    &+\sum_{\langle i,j\rangle} J\Delta S_i^z S_j^z - D\sum_{i} (S_i^z)^2 - g_c \mu_B B\sum_i S_i^z.
\end{align}
We note that the firs term (first line), which correspond to one-magnon process cannot survive from by the projection $P_i=\ket{1_i}\bra{1_i} + \ket{\bar 1_i}\bra{\bar 1_i}$ for low-energy space ($P_iS_i^{\pm} P_i=0$). Consequently, we perform the Schrieffer–Wolff transformation with perturbation $V=J[(1-\eta\Delta S^z_i S^z_j)(S_i^+S_j^-+S_i^-S_j^+)-(S_i^+S_j^-+S_i^-S_j^+)\eta\Delta S^z_i S^z_j]/2$. In contrast to the previous section where $V$ consisted of only two terms [$V=J(S_i^+S_j^-+S_i^-S_j^+)/2$], here $V$ contains six terms, leading to 36 possible contributions to the matrix element product $V_{mk}V_{kn}$. Fortunately, the four additional terms involve only $S^z$ operator, which is diagonal in the spin basis (does not change the spin state). Most importantly, many of these terms vanish because the $S^z$ operator annihilates the only allowed intermediate virtual state $\ket{k}=\ket{0,0}$, as demonstrated previously. Overall, we obtain the effective terms from the perturbation $V$ as
\begin{align}\label{seq:V}
    -\frac{J^2(1-\eta^2\Delta^2)}{2D} |1,\bar 1\rangle \langle \bar 1,1| -\frac{J^2(1-\eta^2\Delta^2)}{2D} |\bar 1, 1\rangle \langle 1,\bar 1| -  \frac{J^2 (1-\eta\Delta )^2}{2D} |1,\bar 1\rangle \langle 1,\bar 1| -\frac{J^2 (1+\eta\Delta)^2 }{2D} |\bar 1, 1\rangle \langle \bar 1, 1|,
\end{align}
where the first two terms can be easily mapped to the biquadratic term
\begin{align}
    -\frac{J^2}{8D}(1-\eta^2\Delta^2) (S^+_iS^+_i S_j^- S_j^- +  S^-_iS^-_i S_j^+ S_j^+). 
\end{align}
The last two terms of Eq.~\eqref{seq:V} need further careful examination. But since they only involve ket and bra states of the same spin state, we can solve the linear the equation $C_0 I_i\otimes I_j +C_1 S_i^z\otimes S_j^z + C_2 I_i\otimes S_j^z  + C_3 S_i^z\otimes I_j $ for the operators of the last two terms. The four coefficients are found to be
\begin{align}
    C_0=-\frac{J^2(1+\eta^2\Delta^2)}{4D},\ C_1=\frac{J^2(1+\eta^2\Delta^2)}{4D},\ C_2= \frac{-J^2\eta\Delta }{2D},\ C_3=\frac{J^2\eta\Delta }{2D}
\end{align}
Note that the $C_0$ term is just a constant term, which can be simply neglected. The $C_2$ and $C_3$ term correspond to a staggered Zeeman field induced by the phonon that couples to the $S^z$ at site $i$ and $j$. Nevertheless, after a summation for each bond of triangular lattice, they cancel out. Therefore, we only need the $C_1$ term. Collecting all the necessary terms, the effective spin Hamiltonian now becomes
\begin{align}
    H_{s}^{\rm eff}=&- \left[\frac{J\eta}{4}+\frac{J^2}{8D}(1-\eta^2\Delta^2)\right] \sum_{\langle i,j \rangle}   (S^+_iS^+_i S_j^- S_j^- +  S^-_iS^-_i S_j^+ S_j^+)- \frac{J\eta}{4}\sum_{\langle i,j\rangle} ( S_i^+S_i^-S_j^-S_j^++S_i^-S_i^+S_j^+S_j^-)\nonumber\\
    &+\sum_{\langle i,j\rangle}\left[ J\Delta + \frac{J^2(1+\eta^2\Delta^2)}{4D}\right] S_i^z S_j^z - D\sum_{i} (S_i^z)^2 - g_c \mu_B B\sum_i S_i^z.
\end{align}

We can now directly apply the projection operator $P_i=|1_i\rangle \langle 1_i| + |\bar 1_i\rangle \langle \bar 1_i|$ to each spin site for the low-energy spin-1/2 space. It's easy to find the following mapping between the effective spin-1/2 space and original spin-1 space
\begin{subequations}
\begin{align}
    &\frac12 P_i S^z_iP_i=\frac12 (|1_i \rangle \langle 1_i| - |\bar 1_i \rangle \langle \bar 1_i| )=\frac12\begin{pmatrix}1 & 0\\0 &-1 \end{pmatrix}=\tau^z_i\\
    &\frac12 P_i S^+_iS^+_i P_i= |1_i \rangle \langle\bar 1_i| = \begin{pmatrix}0 & 1\\0 & 0 \end{pmatrix} =\tau^+_i = (\tau^x_I+\mathrm{i}\tau^y_i)\\
    &\frac12 P_i S^-_iS^-_i P_i= |\bar 1_i \rangle \langle 1_i| = \begin{pmatrix}0 & 0\\1 & 0 \end{pmatrix} =\tau^-_i = (\tau^x_i-\mathrm{i}\tau^y_i)\\
    &\frac12 P_i S^+_i S^-_i P_i= |1_i\rangle \langle 1_i| =\begin{pmatrix}1 & 0\\0 & 0\end{pmatrix}=\tau^0_i+\tau^z_i\\
    &\frac12 P_i S^-_i S^+_i P_i= |\bar 1_i\rangle \langle \bar 1_i| =\begin{pmatrix}0 & 0\\0 & 1\end{pmatrix}=\tau^0_i-\tau^z_i
\end{align}
\end{subequations}
Then, the effective spin Hamiltonian takes the following final form:
\begin{align}\label{seq:H_eff_SW}
    H_{s}^{\rm eff}=-\frac{J^2}{D}\sum_{\langle i,j\rangle } \left[(1+\frac{2D\eta}{J} - \eta^2 \Delta^2 )(\tau^x_i\tau^x_j + \tau^y_i\tau^y_j)- (1+\frac{2D(2\Delta -\eta)}{J} + \eta^2\Delta^2 )\tau^z_i \tau^z_j\right] - 2g_c \mu_B B\sum_i \tau_i^z,
\end{align}
which corresponds to the Eq.(6) of the main text after the substitution $g_c\mu_BB \to B$ (which changes the unit of magnetic field from Tesla to the energy unit Joule).

\section{Estimation for $\eta$}
Here, we discuss and present an order-of-magnitude estimation for $\eta$. The effective mass of the vibrating atom can be set to $M \simeq 60u\simeq 10^{-25}\,\mathrm{kg}$, which is representative of a $3d$ transition-metal ion. We adopt a low intrinsic vibration frequency $ \omega_0  \simeq 62.8\ \mathrm{GHz}$ for the acoustic phonon.  As for the exchange-length sensitivity, we take a conservative range $\gamma\simeq 0.1\text{--}0.5\,\text{\AA}^{-1}$. Putting these values together into the expression $\eta=J\gamma^2/k_E=J\gamma^2/M\omega_0^2$, we obtain
\begin{equation}
\eta \simeq 0.04\text{--}1 \qquad (\gamma=0.1\text{--}0.5~\text{\AA}^{-1}).
\end{equation}
It is worth noting that for a large $\omega_0$ in the sub-THz range, $\eta$ could be significantly suppressed to zero due to the $1/\omega_0^2$ factor in its expression. Furthermore, a large mismatch between magnon and phonon energy scales results in a scenario where mode crossings appear only in the immediate vicinity of the $\Gamma$ point, rendering the phonon probe ineffective. While a large exchange $J$ could address both the energy mismatch and the small $\eta$ issues, it would necessitate an impractically large critical field for the SN phase. On the other hand, although a large $\gamma$ could enhance $\eta$ via the $\gamma^2$ factor in its expression, it fails to bridge the energy gap between magnons and phonons. Therefore, in our numerical calculations, we have adopted a compromise for the energy scales of low-energy acoustic phonon ($\omega_0=62.8 \ \mathrm{GHz}$) and magnon ($\omega_m\equiv J/\hbar=151.7 \ \mathrm{GHz}$) to avoid further parameter tuning. In the phase diagram, the $y$ axis is plotted up to the upper limit $\eta=1$ to provide a comprehensive theoretical overview. We emphasis that, other choices of the $\omega_0,\omega_m,\eta$ do not qualitatively change the conclusion and results of this work.

\section{Spin-nematic phase}
The free phonon Hamiltonian $H_p$ after transforming to the reciprocal space reads
\begin{align}
    H_p=\frac{1}{2M} \sum_k \vec{p}_{k}^\dag I \vec{p}_{k} + 
\frac{1}{2}\sum_k \vec{u}_{k}^\dag \mathcal{D}(k)\vec{u}_k
\end{align}
where $\vec{p}_k=(p^x_{\bm{k}},p^y_{\bm{k}})^T$ and $\mathcal{D}$ is the dynamical matrix in the basis $\vec{u}_k=(u^x_{\bm{k}},u^y_{\bm{k}})^T$. The dynamical matrix $\mathcal{D}$ of the phonon in the triangular lattice of the SN phase is simply a $2\times 2$ matrix, which under the basis of $(u_{\bm{k}}^x,u_{-\bm{k}}^y)$ reads
\begin{align}
    \mathcal{D}(\bm{k})= M\omega_0^2
\begin{pmatrix}
3-2\cos(k_x)-\cos(\frac{k_x}{2})\cos(\frac{\sqrt{3}k_y}{2}) & \sqrt{3}\sin(\frac{k_x}{2})\sin(\frac{\sqrt{3}k_y}{2})\\
\sqrt{3}\sin(\frac{k_x}{2})\sin(\frac{\sqrt{3}k_y}{2}) & 3-3\cos(\frac{k_x}{2})\cos(\frac{\sqrt{3}k_y}{2})
\end{pmatrix}
\end{align}

The phonon operator is defined as $b_{i,k}=\sqrt{\frac{M\omega_{i,k}}{2\hbar}}(U_{i,k}+\frac{i}{M\omega_{i,k}}P_{i,k})$ where $\vec{U}_k = V^\dagger_k \vec{u}_{k}$ and $\vec{P}_k = V^\dagger_k \vec{p}_{k}$ are related to the $\vec{p}_k $ and $\vec{u}_k $ through a unitary transformation $V$ that diagonalizes the dynamical matrix as $V^\dagger \mathcal{D}(\bm{k}) V=M\omega_0^2 {\rm diag}[d_1,d_2]$. The eigenfrequencies of phonon are given by $\omega_{i,\bm{k}}=\omega_0 \sqrt{d_i} $. In the particle-hole-symmetric Nambu basis $\psi_p=(b_{1,\bm{k}},b_{2,\bm{k}},b_{1,-\bm{k}}^\dagger,b^\dagger_{2,-\bm{k}})^T$, the free phonon Hamiltonian can be written as $H_p=\frac{1}{2}\sum_{\bm{k}} \psi^\dagger_{p} \mathcal{H}_p\psi_{p}$, where the kernel $\mathcal{H}_p$ is
\begin{align}
\mathcal{H}_p=\hbar\begin{pmatrix}\omega_{1,\bm{k}} & 0 & 0 & 0\\0 & \omega_{2,\bm{k}} & 0 & 0\\
0 & 0 & \omega_{1,-\bm{k}} & 0\\
0 & 0 & 0 & \omega_{2,-\bm{k}}
\end{pmatrix}.
\end{align}

Regarding the free magnon Hamiltonian $H_s$, we first apply a Schrieffer–Wolff transformation to project the system onto the low-energy subspace. The resulting Hamiltonian retains the same form of Eq.~\eqref{seq:H_eff_SW} but with $\eta=0$. We then perform a Holstein-Primakoff transformation (see below). In reciprocal space, the free magnon Hamiltonian for the SN ground state can be expressed in the Nambu basis $\psi_m=(a_{\bm{k}}, a^\dagger_{-\bm{k}})^T$ as $H_s=\frac12 \sum_{\bm{k}} \psi_m^\dagger \mathcal{H}_s\psi_m$, where $a_{\bm{k}}^\dag (a_{\bm{k}})$ is the magnon creation (annihilation) operator, and the kernel $\mathcal{H}_s$ takes the form:
\begin{align}
    \mathcal{H}_s=\begin{pmatrix}
m_0(\bm{k}) & \Sigma(\bm{k})\\
\Sigma^\dagger (\bm{k})  & m_0(-\bm{k})
\end{pmatrix}.
\end{align}
The off-diagonal term $\Sigma(\bm{k})$ involves a summation over the six nearest-neighbor bonds $\bm{\delta}$ of the triangular lattice, given by $\Sigma=m_4\sum\delta e^{i\bm{k}\cdot \bm{\delta} }$. The diagonal elements are defined by $m_0(k)=m_1 + m_2(\bm{k})+m_3$. Detailed expressions for $m_{1,2,3,4}$ are provided below. We note that dangling terms (proportional to $a_{\bm{k}}$ or $a^\dagger_{\bm{k}}$ alone) may appear in $H_s$; their presence signals an instability in the assumed SN ground state. The coefficient of such linear terms is $ B\sin\phi-3J^2 \sin2\phi(1+\frac{2\Delta D}{J})/2D$. Requiring this coefficient to vanish yields two solutions. The first one, $\phi= 0$, is the trivial solution corresponding to the FP phase. The second one, $\phi= \arccos[2 B D /3J^2(2+\frac{4\Delta D}{J})]$, is the non-trivial solution with lower energy for $B_{c,2}<B<B_{c,3}$, which corresponds to the SN phase. The spin configuration defined by this $\phi$ matches exactly with the one obtained by minimizing the energy in the SN phase at $\eta=0$ [Fig.~\ref{fig:lattice_phase}(b)]. The existence of this non-trivial solution imposes an upper bound on the magnetic field in the SN phase $B_{max}=3J^2(2+\frac{4\Delta D}{J})/2 D $, which is found to coincide with the boundary $B_{c,3}$ in the phase diagram [Fig.1(b) of main text]. Consequently, for a large magnetic fields exceeding the limit ($B>B_{max}=B_{c,3}$), only the ferromagnetic FP phase ($\phi= 0$ remains stable, consistent with the phase diagram.

As for the magnon-phonon coupling $H_{me}$, it does not explicitly depend on the magnetic field. Therefore, the coefficient of magnonic dangling term in $H_{me}$ does not completely vanish as in $H_s$. Therefore, the lowest order in $H_{me}$ becomes bilinear with respect to magnon and phonon operators:
\begin{align}
    H_{me} = \sum_{k} \Pi_i(\bm{k})(b_{i,\bm{k}}+b_{i,-\bm{k}}^\dagger)(a_{-\bm{k}}  +a_{\bm{k}}^\dagger )
\end{align}
The repeated subscript $i$ are assumed to be summed. The detailed expression of $\Pi_i$ can be found below [Eq.~\eqref{seq:Pi}].

In the SN phase, the spins in the effective spin-1/2 space have an ordering wavevector at $\Gamma$, which means the smallest unit cell contains one spin site. As shown in the main text, we parametrize the spin axis with a polar angle $\phi$ as $\hat{\tau}_A=\hat{\tau}_B=\hat{\tau}_C=(\sin\phi,0,\cos\phi)$. We next adopt the Holstein–Primakoff transformation~\cite{Holstein_1940_PR}, which is a Hermitian and exact representation for the spin operators:
\begin{align}\label{seq:HP}
    S_i^\dag =\sqrt{2S} \sqrt{1-\frac{a_i^\dag a_i}{2S}} a_i,\quad S_i^\dag =\sqrt{2S} a_i^\dag \sqrt{1-\frac{a_i^\dag a_i}{2S}},\quad S_i^z=S-a_i^\dag a_i 
\end{align}
with the constraint $0 \leq a_i^\dagger a_i \leq 2S$ enforced by $-S\leq S^z\leq S$. The tricky part lies in the square root of $a_i^\dag a_i $. Usually, one adopts the Taylor expansion in the large $S$ limit. However, it's often used also for the case of $S=1/2$. The justification of this approach comes from the constraint $a_i^\dag a_i \leq 1 $ for $S=1/2$. More rigorously, the Taylor expansion should be replaced with an exact summation~\cite{Gregory_2020_PRR}, which to the lowest order of magnon operator (ignoring magnon-magnon interaction), is the same as the one obtained from Taylor expansion with higher order terms neglected. Thus, around local spin axis, we are still allowed to use Eq.~\eqref{seq:HP} for the effective spin-1/2 model. After rotating back to the global coordinates, we have the following Holstein–Primakoff transformation
\begin{align}\label{seq:tau_HP}
    \tau_{i}^x= \cos\theta_i (a_i+a_i^\dag )/2 + \sin\theta_i (\frac12 -a_i^\dag a_i),\quad \tau_{i}^y= (a_i-a_i^\dag )/(2i),\quad \tau_{i}^z= \cos\theta_i (\frac12 -a_i^\dag a_i)-\sin\theta_i(a_i+a_i^\dag )/2 
\end{align}
with $\theta_i=\phi$ for all the spin sites in the SN phase as we have mentioned. Then, after a tedious calculation, the free magnon Hamiltonian in the reciprocal space is found to be
\begin{align}
    H_s=\frac12 \sum_{\bm{k}} \psi_m^\dag 
\begin{pmatrix}
m_0(\bm{k}) & 2m_4\sum_{\bm{\delta}} e^{i\bm{k}\cdot \bm{\delta} } \\
2m_4\sum_{\bm{\delta}} e^{-i\bm{k}\cdot \bm{\delta} }  & m_0(-\bm{k})
\end{pmatrix}
\psi_m,
\end{align}
where the Nambu basis is $\psi_{m}=(a_{\bm{k}},a_{-\bm{k}}^\dag)$. The $m_i,\ i\in (1,2,3,5) $ terms involved in the Hamiltonian are given by
\begin{align*}
    m_0(k) & =m_1+m_2+m_3\\
    m_1 &=-\frac{12 J^2 }{D}(\frac{\cos2\phi }{4} + \frac{\Delta D}{J}\cos^2\phi)\\
    m_2 &=-\frac{J^2 }{D}[\frac{\cos2\phi+1}{4}-\frac{\Delta D}{J}\sin^2\phi] \sum_{\bm{\delta}}\cos(\bm{k}\cdot \bm{\delta} )\\
    m_3 &= 2g_c \mu_B B \cos\phi\\
    m_4 &= -\frac{J^2 }{2D}[\frac{\cos2\phi-1}{4}-\frac{\Delta D}{J}\sin^2\phi]
\end{align*}
where the summation for dimensionless $\bm{\delta}$ runs over the six nearest-neighbor bond direction of triangular lattice. The linear terms (dangling terms) of free magnon Hamiltonian must vanish for a stable ground state, which enforces the following condition
\begin{align}
    g_c \mu_B B\sin\phi-\frac{3J^2 \sin2\phi}{4D}(2+\frac{4\Delta D}{J})=0,
\end{align}
whose nontrivial solution determines the ground state of $\bm{\tau}$ in the SN phase. 

The total Hamiltonian $H=H_s+H_p +H_{me}$ can then be written in the enlarged Nambu basis $\psi =(b_{1,\bm{k}},b_{2,\bm{k}},b_{1,-\bm{k}}^\dagger,b^\dagger_{2,-\bm{k}},a_{\bm{k}}, a^\dagger_{-\bm{k}})^T$ as $H=\frac{1}{2}\sum_{\bm{k}}\psi^\dagger  \mathcal{H}^{SN} \psi$, where the kernel can be written in the block matrix form
\begin{align}
    \mathcal{H}^{SN}=
    \begin{pmatrix}
        \mathcal{H}_p(\bm{k}) & \Pi^\dagger(\bm{k}) \\
        \Pi(\bm{k}) & \mathcal{H}_s(\bm{k})
    \end{pmatrix}
\end{align}
Here, the phonon kernel has a diagonal form $\mathcal{H}_p={\rm diag}[\omega_{1,\bm{k}},\omega_{2,\bm{k}},\omega_{1,-\bm{k}},\omega_{2,-\bm{k}}]$. The off-diagonal block $\Pi(\bm{k})=\begin{pmatrix}
    \Pi_1 & \Pi_2 & \Pi_1 & \Pi_2\\
    \Pi_1 & \Pi_2 & \Pi_1 & \Pi_2
\end{pmatrix}$
comes from the $H_{me}$, and the matrix element $\Pi_i(\bm{k})$ is
\begin{align}\label{seq:Pi}
    \Pi_j(\bm{k})=-i[d_{j,\bm{k}}]^{-1/4}\sqrt{\frac{\hbar \omega_0 \eta}{2J }}\frac{J^2\sin2\phi}{8 D}(2+\frac{4\Delta D}{J}) \sum_{\bm{\delta}} \sin(\bm{k}\cdot\bm{\delta}) \delta_i V_{ij}
\end{align}

\section{Spin-nematic-supersolid phase}
\begin{figure}[!h]
    \centering
    \includegraphics[width=0.5\linewidth]{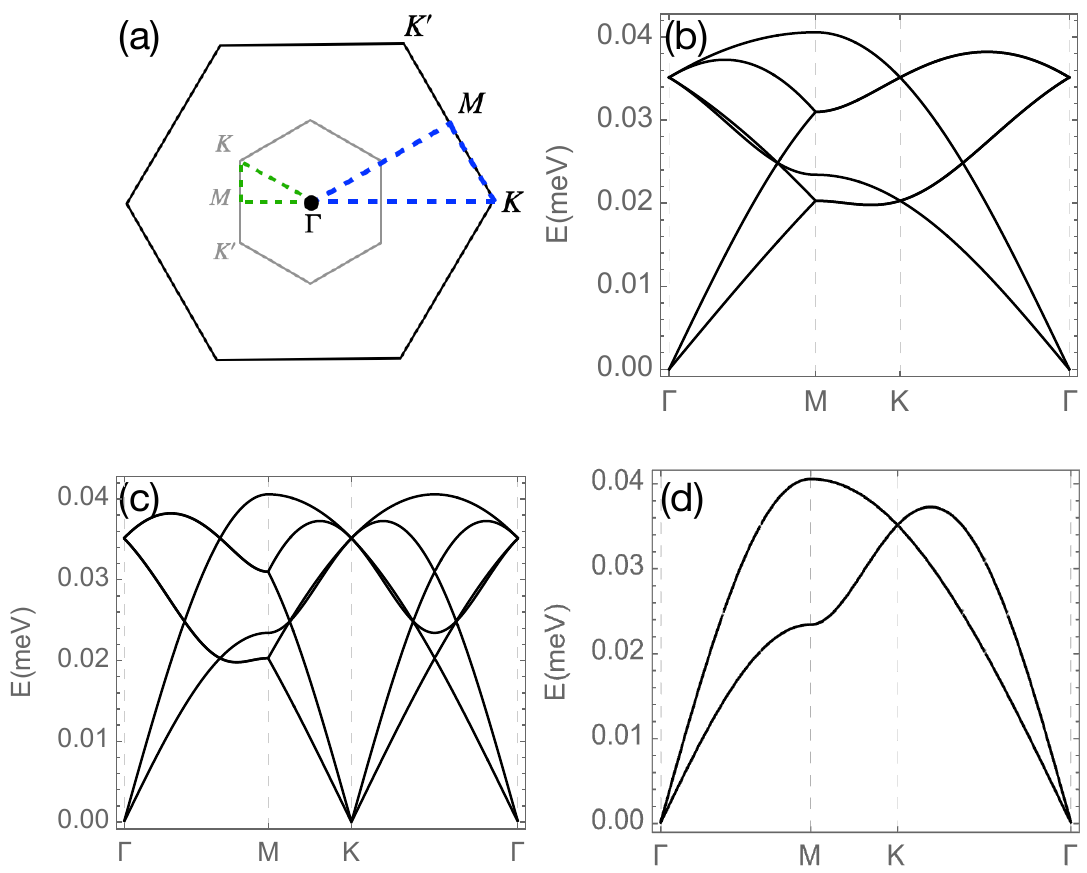}
    \caption{(a) The original BZ (black solid lines) of the triangular lattice and the reduced BZ (gray solid line) corresponding to the enlarged unit cell in the SNS phase. (b) Free phonon band structure along the green dashed path. (c) Free phonon band structure along the blue dashed path. (d) Same as (c), but with the fictitious optical modes removed by projection $P^{ac}$.}
    \label{fig:BZ}
\end{figure}

In SNS phase, we enlarge the unit cell for phonon to be the same as the magnon. This makes the dynamical matrix $\mathcal{D}$ to be a $6 \times 6$ matrix with involved expression. The dynamical matrix $\mathcal{D}$ can be read off from the expression for free phonon Hamiltonian in the reciprocal space
\begin{align}
    H_p= \sum_{\bm{k},\alpha\in A,B,C} \frac{\bm{p}_{\alpha,\bm{k}}^2}{2M} + \frac{M\omega_0^2}{2} \sum_{\bm{k},\alpha\in A,B,C} \sum_{\bm{\delta}^{\alpha\beta }}^{\beta\neq\alpha} (\bm{\delta}^{\alpha\beta}\cdot \bm{u}_{\alpha,\bm{k}}^\dag)(\bm{\delta}^{\alpha\beta}\cdot \bm{u}_{\alpha,\bm{k}})+ (\bm{\delta}^{\alpha\beta}\cdot \bm{u}_{\beta,\bm{k}}^\dag)(\bm{\delta}^{\alpha\beta}\cdot \bm{u}_{\beta,\bm{k}})
\end{align}
The particle-hole-symmetric Nambu basis for the phonon in the enlarge unit cell is  $\psi_p = \psi^-_p \oplus \psi^+_p $ with $\psi^-_p=(b_{1,\bm{k}}, \cdots, b_{6,\bm{k}})^T$ and $\psi^+_p=(b^\dag_{1,-\bm{k}}, \cdots, b_{6,-\bm{k}}^\dag)^T$. The free phonon Hamiltonian then reads $H_p=\frac{1}{2}\sum_{\bm{k}} \psi^\dagger_{p} \mathcal{H}_p\psi_{p}$, where the kernel $\mathcal{H}_p$ is
\begin{align}
\mathcal{H}_p=\hbar\ {\rm diag}[\omega_{1,\bm{k}},\cdots,\omega_{6,\bm{k}},\omega_{1,-\bm{k}}\cdots,\omega_{6,-\bm{k}}].
\end{align}
Here, the eigenfrequency $\omega_{i}$ can be obtained by unitary diagonalizing the $6\times6$ dynamical matrix $\mathcal{D}$.  In Fig.~\ref{fig:BZ}(b), we plot the free phonon band spectra along the $\Gamma$-$M$-$K$-$\Gamma$ path of the reduced BZ [green dashed line in Fig.~\ref{fig:BZ}(a)]. While in Fig.~\ref{fig:BZ}(c), free phonon band spectra is plotted along the same $\Gamma$-$M$-$K$-$\Gamma$ path but of the original BZ [blue dashed line in Fig.~\ref{fig:BZ}(a)]. Note that in both cases, we have six bands in total. By projecting out the fictitious optical modes (see next section) in the original BZ, we obtain Fig.~\ref{fig:BZ}(c), which is in full agreement of the free phonon spectra (blue lines of Fig.2 in the main text). The band spectra of the main text are plotted along the original BZ for phonon [Blue dashed lines].

As for the free magnon Hamiltonian $H_s$, we use the basis $\psi_m = \psi^-_m \oplus \psi^+_m $ with $\psi^-_m=(a_{A,\bm{k}}, a_{B,\bm{k}}, a_{C,\bm{k}})^T$ and $\psi^+_m=(a^\dag_{A,-\bm{k}}, a^\dag_{B,-\bm{k}}, a^\dag_{C,-\bm{k}})^T$. Here, $a^\dag_{\alpha}$ is the magnon creation operator in the $\alpha\in (A,B,C)$ sublattice. We have three bands (three sublattice) for magnon in total. We follow the same procedure as in the SN phase and the free magnon Hamiltonian $H_s$ can be written as $ H_s=\frac12 \sum_{\bm{k}} \psi_m^\dagger \mathcal{H}_s\psi_m$, where the expression of the kernel $\mathcal{H}_s$ becomes very involved and the full expression is provided in Eq.~\eqref{seq:Hs_SNS}. By writing the Hamiltonian $H_s$ in such quadratic form we have assumed that the dangling terms vanish for a stable SNS ground state. The vanishing condition contains two transcendental equations for the angle $\theta_1$ and $\theta_2$ [see Eq.~\eqref{seq:transcendental} below]. The first solution gives $\theta_1=0,\theta_2=\pi$, which is the trivial one and corresponds to the UUD phase. The second solution is the nontrivial one and has the lower energy for $B<B_{c,1}$, which corresponds to the SNS phase. Again, we found that such nontrivial solution exactly matches the one obtained from minimizing the energy in the SNS phase of Fig.1(b) in the main text. Additionally, the existence of the nontrivial solution  (SNS phase) imposes an upper bound $B_{max}$ of the magnetic field for the SNS phase, which is found to exactly match the threshold $B_{max}=B_{c,1}$ in the phase diagram of the main text.

The SNS phase is characterized by $\hat{\bm{\tau}}_A=\hat{\bm{\tau}}_B=(\sin\theta_1,0,\cos\theta_1),\  \hat{\bm{\tau}}_C=(\sin\theta_2,0,\cos\theta_2)$ with ordering wavevector at $K$. Utilizing Eq.~\eqref{seq:tau_HP} and after a tedious calculation, the kernel $\mathcal{H}_s$ of free magnon Hamiltonian is found be
\begin{align}\label{seq:Hs_SNS}
\mathcal{H}_s=
\begin{pmatrix}
d_{1} & m_1^+ f_{AB} & m_{12}^+ f_{AC} & 0 & m_1^- f_{AB} & m_{12}^-f_{AC}\\
m_1^+ f^*_{AB} & d_{1} & m_{12}^+ f_{BC} & m_1^- f^*_{AB} & 0 & m_{12}^- f_{BC}\\
m_{12}^+ f^*_{AC} & m_{12}^+ f^*_{BC} & d_{2} & m_{12}^- f^*_{AC} & m_{12}^- f^*_{BC} & 0\\
0 & m_1^- f_{AB} & m_{12}^- f_{AC} & d_{1} & m_1^+ f_{AB} & m_{12}^+ f_{AC}\\
m_1^- f^*_{AB} & 0 & m_{12}^- f_{BC} & m_1^+ f^*_{AB} & d_1 & m_{12}^+ f_{BC} \\
m_{12}^- f^*_{AC} & m_{12}^- f^*_{BC} & 0 & m_{12}^+ f^*_{AC} & m_{12}^+ f^*_{BC}  & d_2\\
\end{pmatrix}
\end{align}
where the $f_{\alpha\beta},\ \alpha,\beta\in (A,B,C)$ terms are given by $f_{\alpha\beta } = -\frac{J^2}{D} \sum_{\delta^{\alpha\beta }} e^{i\bm{k}\cdot \bm{\delta}^{\alpha\beta }}$. The summation $\sum_{\delta^{\alpha\beta }}$ involves three bond directions that connect the nearest-neighboring sites $\alpha$ and $\beta$ ($\alpha\neq\beta$). The coefficient in the kernel $\mathcal{H}_s$ are
\begin{align*}
    &m_1^{\pm}=\frac{\cos(2\theta_1)\pm 1}{4} - \frac{\sin^2\theta_1 \Delta D}{J}\\
    &m_{12}^{\pm}=[\frac{\cos(\theta_1+\theta_2)\pm 1}{4} - \frac{\Delta D}{J} \sin\theta_1 \sin\theta_2  ]\\
    &d_1= -\frac{J^2}{D}[n_1+n_{12}]+2g_c\mu_B B\cos\theta_1 \\
    &d_2= -2\frac{J^2}{D}n_{12} +2g _c\mu_B B\cos\theta_2\\
    &n_1=3[\frac{\cos 2\theta_{1}}{2} + 2\cos^2\theta_1 \frac{\Delta D}{J}]\\
    &n_{12}=3[\frac{\cos(\theta_1+\theta_2)}{2} + \frac{2\Delta D}{J} \cos\theta_1 \cos\theta_2  ]
\end{align*}
Again, the linear terms (dangling terms) of free magnon Hamiltonian must vanish for a stable ground state, which enforces the following condition
\begin{subequations}\label{seq:transcendental}
\begin{align}
    &-\frac{3J^2}{D}\left[\frac{\sin(\theta_1+\theta_2)+\sin2\theta_1}{4} +\frac{\sin\theta_1 \Delta D}{J}(\cos\theta_2 + \cos\theta_1)\right]+g_c \mu_B B\sin\theta_1 =0\\
    &-\frac{3J^2}{D}\left[\frac{\sin(\theta_1 +\theta_2)}{2}+\frac{2\cos\theta_1 \sin\theta_2 \Delta D}{J}\right]+g_c \mu_B B \sin\theta_2=0
\end{align}
\end{subequations}
The nontrivial solution of above two transcendental equation determines the ground state of $\bm{\tau}$ in the SNS phase.

The magnon-phonon coupling terms have three contribution from the three sublattice $H_{me}=\sum_{\alpha\in(A,B,C)}H_{me}^\alpha$ with $H_{me}^\alpha$ given by
\begin{align}\label{seq:H_me_SNS}
    H^{\alpha}_{me}= &\sum_{k}[d_{j}]^{-\frac14} [G_\alpha]_i[V]_{ij} (b_{j,k}+b_{j,-k}^\dag) (a_{\alpha,-k}+a_{\alpha,k}^\dag),
\end{align}
where the matrix $V$ diagonalizes the dynamical matrix $\mathcal{D}$ of the phonon as $V^\dagger \mathcal{D}(\bm{k}) V=M\omega_0^2 {\rm diag}[d_1,\cdots,d_6]$. The vectors $G_\alpha$ are 
\begin{align*}
&\bm{G}_A= -\frac{J^2 }{D} \sqrt{\frac{\hbar \omega_0 \eta}{2J }} \begin{pmatrix}
0, &0,&
sp_1 g_{AB}^x, & sp_1 g_{AB}^y,&
sp_2 g_{AC}^x, & sp_2 g_{AC}^y
\end{pmatrix}\\
&\bm{G}_B= \frac{J^2 }{D} \sqrt{\frac{\hbar \omega_0 \eta}{2J }} \begin{pmatrix}
sp_1 [g_{AB}^*]^x, & sp_1 [g_{AB}^*]^y,&
0,& 0, & sp_2 g_{BC}^x, & sp_2 g_{BC}^y
\end{pmatrix}\\
&\bm{G}_C= \frac{J^2 }{D} \sqrt{\frac{\hbar \omega_0 \eta}{2J }} \begin{pmatrix}
sp_3 [g_{AC}^*]^x, & sp_3 [g_{AC}^*]^y,&
sp_3 [g_{BC}^*]^x, & sp_3 [g_{BC}^*]^y, &
0,& 0 
\end{pmatrix},
\end{align*}
where the vectors of stricture factor are $\bm{g}_{\alpha\beta} = \sum_{\delta^{\alpha\beta}} e^{i\bm{k}\cdot \bm{\delta}^{\alpha\beta}}\bm{\delta}^{\alpha\beta}$. The summation involves three bond directions that connect the nearest-neighboring sites $\alpha$ and $\beta$ ($\alpha\neq\beta$). The coefficients are appearing in $\bm{G}_\alpha$ are
\begin{align*}
    &sp_1=(2+\frac{4\Delta D}{J})\frac{\sin2\theta_{1}}{8}\\
    &sp_2=[\frac{\cos\theta_1 \sin\theta_2}{4}+\frac{\sin\theta_1 \cos\theta_2}{4}(1+\frac{4\Delta D}{J})]\\
    &sp_3=[\frac{\sin\theta_1 \cos\theta_2}{4}+\frac{\cos\theta_1 \sin\theta_2}{4}(1+\frac{4\Delta D}{J})]
\end{align*}

The total Hamiltonian $H=H_s+H_p +H_{me}$ can then be written in the enlarged Nambu basis $\psi = \psi_p \oplus \psi_m$ as $H=\frac{1}{2}\sum_{\bm{k}}\psi^\dagger  \mathcal{H}^{SNS} \psi$, where the kernel can be written in the block matrix form
\begin{align}\label{seq:H_tot_SNS}
    \mathcal{H}^{SNS}=
    \begin{pmatrix}
        \mathcal{H}_p(\bm{k}) & \Pi^\dagger(\bm{k}) \\
        \Pi(\bm{k}) & \mathcal{H}_s(\bm{k})
    \end{pmatrix}
\end{align}
Here, the off diagonal block $\Pi(\bm{k})$ is a $6\times12$ matrix, whose elements can be easily read off from the Eq.~\eqref{seq:H_me_SNS}. To retrieve the particle-hole symmetry, we first transform our enlarged Nambu basis  $\psi = \psi_p \oplus \psi_m$ to a particle-hole symmetric basis $\Psi = \Lambda \psi$, which results in a particle-hole symmetric Hamiltonian $\mathcal{H}^{SNS}_{p\text{-}h} =(\Lambda^{-1})^{\dag} \mathcal{H}^{SNS} \Lambda^{-1}$. The generalized eigenvalue problem becomes $\mathcal{H}^{SNS}_{p\text{-}h} Q=gQ \bar{\varepsilon}$ with the metric $g=\sigma_z\otimes I_{9\times 9}$ and $\bar{\varepsilon} = g \varepsilon$.

\section{Projection operators}
The projection operator for the acoustic phonon comes naturally from the vibration characteristic at long-wave-length limit ($\bm{k\to 0}$), where the uniform displacement of all sites corresponds to the lattice translational symmetry. Consequently, in the basis of $\psi_u=(u_{A}^x, u_A^y, u_B^x, u_B^y, u_C^x, u_C^y)$, the normalized projection operator $P^u$ for the acoustic mode has the form $P_u=P_u^x (P_u^x)^\dag + P_u^y (P_u^y)^\dag $ with $P_u^x=(1,0,1,0,1,0)^T/\sqrt{3}$ and $P_u^y=(0,1,0,1,0,1)^T/\sqrt{3}$. $P_u^x$ and $P_u^y$ correspond to the uniform vibration along $x$ and $y$ directions, respectively. Taking the expression for $P_u^x$ and $P_u^y$ into $P_u$, we obtain
\begin{align}
    P_u=\frac13\begin{pmatrix}
        1 & 0 & 1 & 0 & 1 & 0\\
        0 & 1 & 0 & 1 & 0 & 1\\
        1 & 0 & 1 & 0 & 1 & 0\\
        0 & 1 & 0 & 1 & 0 & 1\\
        1 & 0 & 1 & 0 & 1 & 0\\
        0 & 1 & 0 & 1 & 0 & 1\\
    \end{pmatrix}
\end{align}
In the particle-hole symmetric Nambu basis $\psi_p=(b_{1,\bm{k}},\cdots,b_{6,\bm{k}}, b_{1,-\bm{k}}^\dag,\cdots, b_{6,-\bm{k}}^\dag)^T$, the projection operator for acoustic mode then becomes $P_p=(V^\dagger p_r V) \oplus (V^\dagger p_r V ) $, where the unitary $V$ comes from changing the basis from $\psi_u$ to $\psi_p$ and the direction sum comes from the hole sector. Note that the Hamiltonian~\eqref{seq:H_tot_SNS} is written in the enlarged Nambu basis $\psi = \psi_p \oplus \psi_m$ with magnon Nambu basis $\psi_m=(a_{A,\bm{k}},a_{B,\bm{k}},a_{C,\bm{k}}, a_{A,-\bm{k}}^\dag,a_{B,-\bm{k}}^\dag,a_{C,-\bm{k}}^\dag)$. To retrieve the particle-hole symmetry for the basis and obtain a generalized eigenvalue problem, we need transform our enlarged Nambu basis  $\psi = \psi_p \oplus \psi_m$ to a particle-hole symmetric basis
\begin{align}
    \Psi\equiv \begin{pmatrix}
    b_{1,k}\\
    \vdots \\
    b_{6,k}\\
    a_{A,k}\\
    a_{B,k}\\
    a_{C,k}\\
    b_{1,-k}^\dag\\
    \vdots \\
    b_{6,-k}^\dag\\
    a_{A,-k}^\dag\\
    a_{B,-k}^\dag\\
    a_{C,-k}^\dag\\
    \end{pmatrix}=\Lambda (\psi_p \oplus \psi_m)
    =\begin{pmatrix}
    I_6 & 0   & 0   & 0   & 0 \\
    0   & 0   & 0   & I_3 & 0 \\
    0   & I_3 & 0   & 0   & 0 \\
    0   & 0   & I_3 & 0   & 0 \\
    0   & 0   & 0   & 0   & I_3
    \end{pmatrix}
    \begin{pmatrix}
    b_{1,k}\\
    \vdots \\
    b_{6,k}\\
    b_{1,-k}^\dag\\
    \vdots \\
    b_{6,-k}^\dag\\
    a_{A,k}\\
    a_{B,k}\\
    a_{C,k}\\
    a_{A,-k}^\dag\\
    a_{B,-k}^\dag\\
    a_{C,-k}^\dag\\
    \end{pmatrix}
\end{align}
Then, in the particle-hole symmetric Nambu $\Psi$, the projection operator for the acoustic phonon $P^{ax}$ takes the final form
\begin{align}
    P^{ac}= (\Lambda^{-1})^\dag \left[(V^\dagger p_r V) \oplus (V^\dagger p_r V ) \oplus 0_{6\times 6} \right] \Lambda^{-1}
\end{align}
Note that, the above operator also projects out the magnon bands. To keep the magnon band and simultaneously remove the fictitious optical phonon bands, we can replace the last $0_{6\times 6}$ matrix with identity matrix $I_{6\times 6}$. The projection operator we constructed works very well where in Fig.~\ref{fig:BZ}(d) we can see that four bands are completely removed and the remaining two bands are exactly the acounstic phonon bands. In Figs.4(c,f) of the main text, each band is plotted with the transparency weight given by $T_n= Q_n^\dag g P^{ax} Q_n$ such that the fictitious optical modes will become invisible (removed). Here, $Q_n$ is the $n$-th column of eigenstate matrix $Q$ for the  particle-hole symmetric Hamiltonian $\mathcal{H}^{SNS}_{p\text{-}h} =(\Lambda^{-1})^{\dag} \mathcal{H}^{SNS} \Lambda^{-1}$.

To visualize the phonon-magnon hybridization, we plot the band spectra with color gradients representing the normalized relative weights of the phonon-like (blue) and magnon-like (red) components for the $n$-th band. The weights are defined as $w_a^n=Q_n^\dagger gP_aQ_n$ and $w_b^n=Q_n^\dagger gP_bQ_n$, where $Q_n$ is the $n$-th column of the transformation (eigenstate) matrix $Q$, and $P_{a/b}$ are the normalized projection operators for the magnon and phonon subspaces, respectively.  In the particle-hole symmetric Nambu $\Psi$, the projection operator for phonon-like $P_b$ and magnon-like $P_a$ components is straightforwardly given by $P_b={\rm diag}[1,\cdots,1,0,0,0,1,\cdots,1,0,0,0]$ and  $P_b={\rm diag}[0,\cdots,0,1,1,1,0,\cdots,0,1,1,1]$. These weights satisfy the normalization condition $w_a^n+w_b^n=Q^\dag_n gQ_n = 1$ (since $P_a+P_b=I$) for the bands in the particle sectors with positive energy. As illustrated in Fig.2 of main text, the bands near the avoided crossing appear green, signifying strong hybridization between magnon and phonon states. We note that while a finite $\eta$ does not open a gap at the $\Gamma$ point, the eigenstates in its vicinity exhibit mixed character (green color) at high magnetic fields [Fig.2(f) of main text], indicating presence of hybridization despite the absence of an avoided crossing gap.

\end{document}